\documentclass[12pt]{article}
\usepackage{epsfig}
\usepackage{amssymb}
\usepackage{amsmath}
\usepackage{amsfonts}
\usepackage{graphicx}
\usepackage{mathrsfs}
\usepackage[dvips]{color}
\usepackage{multirow}

\newcommand{\R}{\mathbb{R}}
\newcommand{\C}{\mathbb{C}}
\newcommand{\Z}{\mathbb{Z}}

\newcommand{\fa}{\mathfrak{a}}
\newcommand{\fb}{\mathfrak{b}}

\newcommand{\fn}{{\,\mathfrak{n}\,}}

\newcommand{\fz}{\mathfrak{z}}

\newcommand{\bH}{\mathbf{H}}

\newcommand{\bM}{\mathbf{M}}
\newcommand{\bN}{\mathbf{N}}

\newcommand{\bS}{\mathbf{S}}

\newcommand{\bU}{\mathbf{U}}

\newcommand{\bsigma}{\mathbf{\sigma}}

\newcommand{\cF}{\mathcal{F}}

\newcommand{\cP}{\mathcal{P}}

\newcommand{\cT}{\mathcal{T}}
\newcommand{\cU}{\mathcal{U}}

\newcommand{\be}{\begin{equation}}
\newcommand{\ee}{\end{equation}}
\newcommand{\bea}{\begin{eqnarray}}
\newcommand{\eea}{\end{eqnarray}}
\newcommand{\nn}{\nonumber}
\newcommand{\kt}{\rangle}
\newcommand{\br}{\langle}

\newcommand{\ed}{\end{document}}

\newcommand{\np}{\newpage}

\newcommand{\bi}{\begin{itemize}}
\newcommand{\ei}{\end{itemize}}

\newcommand{\bce}{\begin{center}}
\newcommand{\ece}{\end{center}}

\newcommand{\sT}{\mathscr{T}}

\newcommand{\RE}{{\rm Re}}
\newcommand{\IM}{{\rm Im}}

\begin{document}

\title{Adiabatic Approximation, Semiclassical Scattering,\\ and Unidirectional Invisibility}

\author{Ali~Mostafazadeh\thanks{E-mail address:
amostafazadeh@ku.edu.tr, Phone: +90 212 338 1462, Fax: +90 212 338
1559}
\\
Department of Mathematics, Ko\c{c} University,\\
34450 Sar{\i}yer, Istanbul, Turkey}
\date{ }
\maketitle

\begin{abstract}

The transfer matrix of a possibly complex and energy-dependent scattering potential can be identified with the $S$-matrix of a two-level time-dependent non-Hermitian Hamiltonian $\bH(\tau)$. We show that the application of the adiabatic approximation to $\bH(\tau)$ corresponds to the semiclassical description of the original scattering problem. In particular, the geometric part of the phase of the evolving eigenvectors of $\bH(\tau)$ gives the pre-exponential factor of the WKB wave functions. We use these observations to give an explicit semiclassical expression for the transfer matrix. This allows for a detailed study of the semiclassical unidirectional reflectionlessness and invisibility. We examine concrete realizations of the latter in the realm of optics.

\medskip

\noindent {Pacs numbers: 03.65.Nk, 03.65.Sq, 03.65.Vf, 42.25.Bs}
\end{abstract}

\maketitle

\section{Introduction}

The discovery of surprising physical phenomena such as lasing at threshold gain \cite{laser1,laser2}, antilasing \cite{antilaser}, $\cP\cT$-symmetric and non-$\cP\cT$-symmetric CPA lasers \cite{CPA-laser,jpa-2012}, and unidirectional invisibility \cite{invisible0,invisible1,invisible2,invisible3}, which are only supported by complex scattering potentials \cite{prl-2009,pra-2013a}, has provided a renewal of interest in the study of complex potential scattering in one dimension \cite{muga}. Among the remarkable outcomes of this study are a dynamical formulation of one-dimensional scattering theory \cite{p114}, which identifies the scattering data with the solutions of a set of dynamical equations, and the surprising observation that the transfer matrix of any scattering potential coincides with the $S$-matrix of an associated non-stationary and non-unity two-level quantum system~\cite{p113}. The purpose of the present article is to use the machinery of the adiabatic approximation to determine the dynamics of the latter system. This provides a semiclassical description of the scattering phenomenon that we can use to develop a semiclassical treatment of spectral singularities and unidirectional invisibility.

The utility of semiclassical or WKB approximation in scattering theory has a long history \cite{FW,austern,Newton,muga}. One-dimensional semiclassical scattering theory has attracted much attention in the study of the mathematical aspects of scattering theory \cite{RC} and led to the development of the phase integral methods \cite{exact-WKB}. It has also served as an important tool in the study of complex absorbing potentials \cite{vibok-riss-manolopoulos,muga}. Our investigation of the one-dimensional semiclassical scattering follows a different path and is particularly suitable for the study of the scattering properties of complex potentials arising in optics.

In the remainder of this section we survey the necessary background material that we use in the rest of the article.

Consider the time-independent Schr\"odinger equation
    \be
	-\psi''(x)+v(x)\psi=k^2\psi(x),
	\label{sch}
	\ee
for a possibly energy-dependent complex scattering potential $v$ defined on the real line. We can describe the scattering properties of $v$ using its transfer matrix \cite{muga,sanchez-soto}. This is a $2\times 2$ matrix that is related to the reflection and transmission amplitudes, $R^{r/l}$ and $T$, according to
    \be
    \bM=\left[\begin{array}{cc}
    \displaystyle T-\frac{R^lR^r}{T} & \displaystyle \frac{R^r}{T} \vspace{.2cm}\\
    \displaystyle -\frac{R^l}{T} & \displaystyle \frac{1}{T}\end{array}\right].
    \label{M}
    \ee
Recall that $T$ and $R^{l/r}$ are the $k$-dependent coefficients determining the asymptotic expression for the scattering solutions $\psi_k^{l/r}$ of (\ref{sch}) according to
    \begin{align}
	&\psi_k^l(x)=\left\{
	\begin{array}{ccc}
	e^{ikx}+R^l e^{-ikx} & {\rm for} & x\to-\infty,\\[6pt]
	T e^{ikx}& {\rm for} & x\to\infty,
	\end{array}\right.&&
	\psi_k^r(x)=\left\{ \begin{array}{ccc}
	T e^{-ikx} & {\rm for} & x\to-\infty,\\[6pt]
	e^{-ikx}+R^r e^{ikx}& {\rm for} & x\to\infty.
	\end{array}\right.
	\label{scatter}
	\end{align}

It is easy to show that the time-independent Schr\"odinger equation (\ref{sch}) is equivalent to the time-dependent Schr\"odinger equation,
    \be
	i\dot\Psi(\tau)=\bH(\tau)\Psi(\tau),
	\label{sch-eq}
	\ee
where $\tau:=kx$, an over-dot represents a derivative with respect to $\tau$,
	\begin{align}
	\Psi(\tau)&:=\frac{1}{2}\left[\begin{array}{c}
	\phi(\tau)-i\dot\phi(\tau)\\
	\phi(\tau)+i\dot\phi(\tau)\end{array}\right],~~~~~~~~~\phi(\tau):=\psi(\tau/k),
	\label{def1}\\
	\bH(\tau)&:=\left[\begin{array}{cc}
	w(\tau)-1 & w(\tau)\\
	-w(\tau) & -w(\tau)+1\end{array}\right]
	=-\bsigma_3+w(\tau)\bN,
	\label{def2}\\
	w(\tau)&:=\frac{v(\tau/k)}{2k^2},~~~~~~~~~
    \bN:=i\bsigma_2+\bsigma_3=
    \left[\begin{array}{cc} 1 & 1 \\-1 & -1\end{array}\right],
	\end{align}
and $\bsigma_i$, with $i=1,2,3$, are Pauli matrices \cite{p113}. The Hamiltonian $\bH(\tau)$ is manifestly time-dependent and non-Hermitian.\footnote{Because $\tau$ is a dimensionless evolution parameter for the two-level quantum system defined by the Hamiltonian (\ref{def2}), we refer to $\tau$ as time.} For real potentials, it is $\sigma_3$-pseudo-Hermitian \cite{p123}, and its exceptional points correspond to the classical turning points of $v$, where $v(x)=k^2$, \cite{p113}.
If we denote the evolution operator for $\bH(\tau)$ by $\bU(\tau,\tau_0)$, i.e., $\bU(\tau,\tau_0):=\sT e^{-i\int_{\tau_0}^{\tau}d\tau'\bH(\tau')}$, where $\sT$ stands for the time-ordering operator, and let
	\be
	U_0(\tau):=e^{i\tau\sigma_3},
	\label{u0=}
	\ee
the $S$-matrix of $\bH(\tau)$ takes the form \cite{weinberg}
    \be
    \bS:=\bU_0(+\infty)^{-1}\bU(+\infty,-\infty)\bU_0(-\infty).
    \label{S}
    \ee
A remarkable property of $\bH(\tau)$ is that its $S$-matrix gives the transfer matrix of $v$, \cite{p113}, i.e.,
    \be
    \bS=\bM.
    \label{S=M}
    \ee
This equation reduces the solution of the scattering problem for the potential $v$ to the determination of the time-evolution operator $\bU(\tau,\tau_0)$ defined by $\bH(\tau)$.

The fact that $\bH(\tau)$ is a time-dependent Hamiltonian operator suggests exploring the consequences of applying adiabatic approximation for the determination of $\bU(\tau,\tau_0)$.

\section{Adiabatic and Semiclassical Approximations}

Adiabatic approximation can be applied to non-Hermitian matrix Hamiltonians \cite{GW} provided that the path in the parameter space of the Hamiltonian, which determines its time-dependence, does not intersect an exceptional point \cite{ep,jmp-2008}. As we noted in Ref.~\cite{p113}, exceptional points of the Hamiltonian $\bH(\tau)$ correspond to classical turning points, where $v(x)=k^2$. This shows that we can safely apply the adiabatic approximation for the cases that $|v(x)|$ is bounded from above and $k$ exceeds this bound appreciably.

In order to apply the adiabatic approximation for the Hamiltonian $\bH(\tau)$, we need to solve the eigenvalue problem for $\bH(\tau)$ and construct a complete orthonormal system \cite{review}, $\{(\Psi_\pm(\tau),\Phi_\pm(\tau))\}$, such that $\Psi_\pm(\tau)$ and $\Phi_\pm(\tau)$ are respectively the eigenvectors of $\bH(\tau)$ and $\bH(\tau)^\dagger$, \cite{GW}. A straightforward calculation with a symmetric choice for the eigenvectors $\Psi_\pm(\tau)$ gives
	\begin{align}
	&E_\pm(\tau):=\pm\fn(\tau), &&
	\Psi_\pm(\tau):=\frac{1}{2}\left[\begin{array}{c}
	1\mp\fn(\tau)\\
	1\pm\fn(\tau)\end{array}\right], &&
    \Phi_\pm(\tau):=\frac{1}{2\fn(\tau)^*}\left[\begin{array}{c}
    \fn(\tau)^*\mp 1\\
    \fn(\tau)^*\pm 1\end{array}\right],
	\label{eg-val-vec}
	\end{align}
where
    \be
    \fn(\tau):=\sqrt{1- 2 w(\tau)}=\sqrt{1-\frac{v(\tau/k)}{k^2}}.
    \ee

Adiabatic approximation asserts that as time $\tau$ progresses the eigenstates of the initial Hamiltonian $\bH(\tau_0)$ evolve into the eigenstates of $\bH(\tau)$. More specifically, we have
    \be
    \Psi_\pm(\tau_0)\to \Psi(\tau)\approx
    e^{i([\delta_\pm(\tau)+\gamma_\pm(\tau)]}\Psi_\pm(\tau)=
    \frac{1}{2}\sqrt{\frac{\fn(\tau_0)}{\fn(\tau)}}\:e^{\mp i\int_{\tau_0}^\tau E_\pm(\tau')d\tau'}
    \left[\begin{array}{c}
	1\mp\fn(\tau)\\
	1\pm\fn(\tau)\end{array}\right],
    \label{adiab1}
    \ee
where
   \bea
    &&\delta_\pm(\tau)=-\int_{\tau_0}^\tau E_\pm(\tau')d\tau',
    \label{d=}\\
    &&\gamma_\pm(\tau)=i\int_{\tau_0}^\tau \br\Phi_\pm(\tau')|d\Psi_\pm(\tau')\kt d\tau'.
    \label{g=}
    \eea
Notice that $e^{i\delta_\pm(\tau)}$ and $e^{i\gamma_\pm(\tau)}$ are respectively the dynamical and geometrical phases \cite{Berry} associated with the initial state vector $\Psi_\pm(\tau_0)$ whenever $\bH(\tau)=\bH(\tau_0)$. For an arbitrary initial state vector $\Psi(\tau_0)$, adiabatic approximation implies
    \be
    \Psi(\tau_0)\to \Psi(\tau)\approx \cU_0(\tau,\tau_0)\Psi_\pm(\tau_0),
    \ee
where $\cU_0(\tau,\tau_0)$ is the adiabatic evolution operator \cite{pra-1997-jmp-1999,book1} for the Hamiltonian $\bH(\tau)$, i.e.,
    \be
    \cU_0(\tau,\tau_0):=\sum_{a=\pm} e^{i([\delta_a(\tau)+\gamma_a(\tau)]}|\Psi_a(\tau)\kt\br\Phi_a(\tau_0)|.
    \label{adia-U}
    \ee

We can easily compute the ``geometric'' factor $e^{i\gamma_\pm(\tau)}$ entering (\ref{adiab1}) and (\ref{adia-U}). In view of (\ref{eg-val-vec}) and (\ref{g=}), it is given by
    \be
    e^{i\gamma_\pm(\tau)}=\sqrt{\frac{\fn(\tau_0)}{\fn(\tau)}}.
    \label{geo=}
    \ee
Similarly, we observe that the condition for the validity of the adiabatic approximation, namely
$\left|\br\Phi_\pm|\frac{d}{d\tau}\Psi_\mp\kt\right/[E_+(\tau)-E_-(\tau)]|\ll 1$, has the explicit form
	\be
	\left| \frac{\dot\fn(\tau)}{4\fn(\tau)^2}\right|\ll 1.
	\label{adi-condi-0}
	\ee
Expressing this condition in terms of the potential, we find $|v'(x)|/8|k^2-v(x)|^{3/2}\ll 1$,
which is often given as the condition for the validity of the semiclassical approximation.

In order to elucidate the relationship between the adiabatic and semiclassical approximations, we identify the right-hand sides of (\ref{def1}) and (\ref{adiab1}) to determine the form of the single-component state vector $\phi(\tau)$ for the adiabatically evolving two-component state vectors (\ref{adiab1}). Because $\psi(x)=\psi(\tau/k)=\phi(\tau)$, this gives
    \bea
    \psi(x)=\psi^{\mp}(x)&:=&\sqrt{\frac{\fn(\tau_0)}{\fn(\tau)}}\:e^{\mp i\int_{\tau_0}^\tau E_\pm(\tau')d\tau'}=N_0 \left[k^2-v(x)\right]^{-1/4} e^{\mp i\int_{x_0}^x
    \sqrt{k^2-v(x')}dx'},
    \label{WKB=}
    \eea
where $N_0:=\left[k^2-v(x_0)\right]^{1/4}$ and $x_0:=\tau_0/k$. The wave functions $\psi^{\pm}$ are precisely the semiclassical solutions of the time-independent Schr\"odinger equation~(\ref{sch}) that are often referred to as the WKB wave functions. Therefore, the two-component formulation of this equation described in Section~1 provides a clear demonstration of the equivalence of the adiabatic and semiclassical approximations. In this context, it is remarkable that the geometric part (\ref{geo=}) of the (complex) phase factor contributing to the expression for the adiabatically evolving two-component state vectors~(\ref{adiab1}) corresponds to the pre-exponential factor of the WKB wave functions (\ref{WKB=}).

\section{Semiclassical Transfer Matrix}

The coincidence of the adiabatic and semiclassical approximations suggests that we express the latter in the form
    \be
    \bU(\tau,\tau_0)\approx\cU_0(\tau,\tau_0),
    \label{adi-sem-app}
    \ee
and use this relation together with (\ref{S}) and (\ref{S=M}) to determine the semiclassically approximate transfer matrix $\bM_{\rm sc}$.

For an infinite-range potential fulfilling the adiabaticity condition~(\ref{adi-condi-0}) for all $\tau\in\R$, Eqs.~(\ref{S}), (\ref{S=M}), (\ref{adia-U}) and (\ref{adi-sem-app}) lead to
    \be
    \bM\approx\bM_{\rm sc}=\left[
    \begin{array}{cc}
    e^{i\vartheta}& 0\\
    0 & e^{-i\vartheta}\end{array}\right],~~~~~~~~~
    \vartheta:=\int_{-\infty}^\infty[\fn(\tau)-1]\,d\tau=
    \int_{-\infty}^\infty[\sqrt{k^2-v(x)}-k]\,dx.
    \label{wkb-infinite}
    \ee
Therefore, in this case, the semiclassical approximation ignores the effect of the (above barrier) reflection, and for a real potential gives a unit transmission coefficient, $|T|^2\approx 1$, in conformity with the prediction of classical mechanics \cite{holstein}.

In the remainder of this section we explore the implications of (\ref{adi-sem-app}) for scattering due to a finite-range potential.

In order to determine the semiclassical transfer matrix for a finite-range potential, first we introduce  $\tau_\pm$ as the real numbers such that $(\tau_-,\tau_+)$ is the largest open interval in $\R$ outside which $w(\tau)$ vanishes identically.  We can then restate Eq.~(\ref{S=M}) in the form  \cite{p113},
    \be
    \bM=\bU_0(\tau_+)^{-1}\bU(\tau_+,\tau_-)\bU_0(\tau_-).
    \ee
Substituting (\ref{adi-sem-app}) on the right-hand side of this equation and employing (\ref{u0=}), we obtain
    \be
    \bM\approx\bM_{\rm sc}:=\left[
    \begin{array}{cc}
    e^{-i(\tau_+-\tau_-)}(\fa_+\cos\delta+i\fb_+\sin\delta)&
    e^{-i(\tau_++\tau_-)}(\fa_-\cos\delta+i\fb_-\sin\delta)\\
    e^{i(\tau_++\tau_-)}(\fa_-\cos\delta-i\fb_-\sin\delta) &
    e^{i(\tau_+-\tau_-)}(\fa_+\cos\delta-i\fb_+\sin\delta)\end{array}\right],
    \label{M-56}
    \ee
where
	\begin{align}
	&\fa_\pm:=\frac{1}{2}\left[\sqrt{\frac{\fn_-}{\fn_+}}\pm
	\sqrt{\frac{\fn_+}{\fn_-}}\right], &&
	\fb_\pm:=\frac{1}{2}\left[\sqrt{\fn_-\fn_+}\pm\frac{1}{\sqrt{\fn_-\fn_+}}\right],
	\label{fas-1}\\
	&\fn_\pm:=\fn(\tau_\pm)=\sqrt{1-\frac{v(x_\pm)}{k^2}}, &&
	\delta:=\delta_-(\tau_+)=\int_{\tau_-}^{\tau_+}\!\! \fn(\tau)\,d\tau=
	\int_{x_-}^{x_+}\!\! \sqrt{k^2-v(x)}\,dx,
	\label{fas-2}
	\end{align}
and $x_\pm:=\tau_\pm/k$.

As a simple example, consider the application of (\ref{M-56}) for the barrier potential:
   	\be
	v(x)=\left\{\begin{array}{cc}
    	\fz & {\rm for}~x\in(0,L),\\
    	0 & {\rm otherwise},\end{array}\right.
	\label{barrier}
	\ee
where $\fz\in\C$ and $L\in\R^+$. For this potential, we have $x_-=0$ and $x_+=L$ which together with (\ref{fas-1}) -- (\ref{barrier})  imply
    \begin{align*}
    &\fn_\pm=\fn=\sqrt{1-\fz/k^2}, &&\fa_+=1, &&\fa_-=0, &&\fb_\pm=\frac{\fn^2\pm 1}{2\fn}.
    \end{align*}
Inserting these equations in (\ref{M-56}) and comparing the result with the exact expression for
the transfer matrix of this potential \cite{pra-2013a}, namely
    \be
    \bM=\left[\begin{array}{cc}
    \left\{\cos(\!\fn k L)+\displaystyle
	\frac{i(\fn^2+1)\sin(\!\fn k L)}{2\fn}\right\}e^{-ik L} &
    \displaystyle\frac{i(\fn^2-1)\sin(\!\fn k L) e^{-ik L}}{2\fn}\\[12pt]
    \displaystyle-\frac{i(\fn^2-1)\sin(\!\fn k L) e^{ik L}}{2\fn} &
    \left\{\cos(\!\fn k L)-
    \displaystyle\frac{i(\fn^2+1)\sin(\!\fn k L)}{2\fn}\right\}e^{ik L}\end{array}\right],
    \nn
    \ee
we find $\bM_{\rm sc}=\bM$. This was to be expected, because for this potential $\dot\fn=0$ and the adiabatic (semiclassical) approximation is exact.

Next, we use (\ref{M}) and (\ref{M-56}) to derive the following explicit form of the semiclassical approximation for reflection and transmission amplitudes.
	\bea
	&&R^l\approx R^l_{\rm sc}:=-e^{2i\tau_-}\left(\frac{\fa_--i\fb_-\tan\delta}{\fa_+-i\fb_+\tan\delta}\right),
	\label{RL-sc}\\[6pt]
	&&R^r\approx R^r_{\rm sc}:=e^{-2i\tau_+}\left(\frac{\fa_-+i\fb_-\tan\delta}{\fa_+-i\fb_+\tan\delta}\right),\label{RR-sc}\\[6pt]
	&&T\approx  T_{\rm sc}:= \frac{e^{-i(\tau_+-\tau_-)}}{\fa_+\cos\delta-i\fb_+\sin\delta}.
	\label{T-sc}
	\eea
We can use these relations to study semiclassical spectral singularities, where $R^{l/r}_{\rm sc}$ and $T_{\rm sc}$ diverge, and semiclassical unidirectional invisibility, where one of $R^l_{\rm sc}$ and
$R^r_{\rm sc}$ vanishes and $T_{\rm sc}=1$. We treat the former in Ref.~\cite{pra-2011b} using a more direct but elaborate approach. We devote the next section for a discussion of the latter.

In the following we respectively use the terms ``left- (right-) reflectionless''  and ``left- (right-) \linebreak invisible'' to mean ``unidirectionally reflectionless from the left (right)'' and ``unidirectionally invisible from the left (right),'' for brevity.

\section{Semiclassical Unidirectional Invisibility}

According to (\ref{RL-sc}) the semiclassical left-reflectionlessness corresponds to the condition
	\be
	\tan\delta=\frac{-i\fa_-}{\fb_-},
	\label{condi-sc1}
	\ee
which, in view of (\ref{fas-1}), is equivalent to
	\be
	e^{2i\delta}=\left(\frac{\fn_--1}{\fn_-+1}\right)\left(\frac{\fn_++1}{\fn_+-1}\right).
	\label{condi-ref-1}
	\ee
An obvious consequence of this relation is
    \be
    \fn_\pm\neq 1.
    \label{condi-01}
    \ee
Substituting (\ref{condi-sc1}) in (\ref{RR-sc}) and (\ref{T-sc}) gives
	\bea
	R^r_{\rm sc}&=&\frac{(\fn_--\fn_+)(\fn_-\fn_+-1)e^{-2i\tau_+}}{\fn_-(\fn_+^2-1)},
	\label{RR-sc-ref}\\
	T_{\rm sc}&=& e^{-i(\tau_+-\tau_-)}\sqrt{\frac{\fn_--\fn_-^{-1}}{\fn_+-\fn_+^{-1}}}.
	\label{T-sc-ref}
	\eea
Therefore, if
	\be
	\fn_-\neq\fn_+\neq\fn_-^{-1},
	\label{condo-zero}
	\ee
and (\ref{condi-ref-1}) holds for a real value of $k$, the potential $v$ possesses semiclassical left-reflectionlessness for this value of $k$. In this case, according to (\ref{RR-sc-ref}) and (\ref{T-sc-ref}), the semiclassical reflection coefficient from the right $|R^r_{\rm sc}|^2$ and the transmission coefficient $|T_{\rm sc}|^2$ only depend on the boundary values of the refractive index, $\fn_\pm$. Surprisingly, they are even independent of the wavenumber $k$. This proves the following interesting result.
    \begin{itemize}
    \item[]\textbf{Theorem}: Let $x_\pm$ and $k_0$ be real numbers such that
    $k_0>0>x_--x_+$, and $v:\R\to\C$ be a finite-range smooth potential with support $[x_-,x_+]$. Suppose that for all $k\geq k_0$ the potential $v$ satisfies (\ref{adi-condi-0}), equivalently
            \be
            \left|\frac{1}{\fn}\frac{d\fn}{d\hat x}\right|\ll 4 k L~~~~~{\rm for}~~~~0\leq\hat x\leq 1,
            \label{sc-condi-375}
            \ee
    where $\fn:=\sqrt{1-v^2/k^2}$ and $\hat x:=(x-x_-)/(x_+-x_-)$. If $v$ is unidirectionally reflectionless for some wavenumber $k_\star\geq k_0$. Then its reflection and transmission coefficients, $|R^{r/l}|^2$ and $|T|^2$, have the same values for any other wavenumber for which it is unidirectionally reflectionless. Moreover, these coefficients are uniquely determined by the boundary values of $v$, i.e., $v(x_\pm)$.
    \end{itemize}

Now, consider the case where in addition to (\ref{condi-ref-1}) and (\ref{condo-zero}), we also have
	\be
	e^{i(\tau_+-\tau_-)}=\sqrt{\frac{\fn_--\fn_-^{-1}}{\fn_+-\fn_+^{-1}}}.
	\label{condi-inv-1}
	\ee
Then $T_{\rm sc}=1$ and $v$ possesses semiclassical left-invisibility. Notice that because $\tau_\pm$ are real parameters, (\ref{condi-inv-1}) is equivalent to
	\bea
	&&\left|\fn_--\fn_-^{-1}\right|=\left|\fn_+-\fn_+^{-1}\right|\neq 0,
	\label{condi-inv-2}\\
	&&\frac{1}{2}\: {\rm arg}\left(\frac{\fn_--\fn_-^{-1}}{\fn_+-\fn_+^{-1}}\right)=
    \tau_+-\tau_--\pi m=kL-\pi m,
	\label{condi-inv-3}
	\eea
where for every $\fa\in\C$, ``${\rm arg}(\fa)$'' stands for the principal argument (phase angle) of $\fa$ that takes values in $(-\pi,\pi]$, $m$ is an integer, and $L:=x_+-x_-=(\tau_+-\tau_-)/k$.

Relation~(\ref{condi-inv-2}) is a constraint on the boundary values of $\fn$ or equivalently the potential $v$.  We can offer a more explicit form of this constraint by introducing $r:=\left|\fn_--\fn_-^{-1}\right|$ and $\varphi_\pm:={\rm arg}(\fn_\pm)$, so that
    \be
    \fn_\pm=|\fn_\pm|e^{i\varphi_\pm}.
    \label{polar-npm}
    \ee
We do this by squaring
$\left|\fn_\pm-\fn_\pm^{-1}\right|$ and using (\ref{condi-inv-2}) to obtain the following quadratic equation for $|\fn_\pm|^2$.
    \be
    |\fn_\pm|^4-2s_\pm|\fn_\pm|^2+1=0,
    \label{eqnnn}
    \ee
where $s_\pm:=\cos(2\varphi_\pm)+ r^2/2$. Eq.~(\ref{eqnnn}) has real and positive solutions provided that $s_\pm\geq 0$ and $s_\pm^2\geq1$. This in turn implies $s_\pm\geq1$. Hence $r$ and $\varphi_\pm$ satisfy
	\be
	\cos(2\varphi_\pm)+ \frac{r^2}{2}\geq1.
	\label{condo-spm}
	\ee
Under this condition, Eq.~(\ref{eqnnn}) implies
	\be
	|\fn_\pm|=\sqrt{s_\pm+\varepsilon_\pm\sqrt{s_\pm^2-1}},
	\label{abs-npm}
	\ee
where
	\[\varepsilon_\pm:={\rm sgn}(|\fn_\pm|-1)=\left\{\begin{array}{cc}
	+&{\rm for}~|\fn_\pm|> 1,\\
	0 &{\rm for}~|\fn_\pm|=1,\\
	-&{\rm for}~|\fn_\pm|< 1.\end{array}\right.\]

Next, we substitute  (\ref{abs-npm}) in (\ref{polar-npm}) and use the result in (\ref{condi-inv-3}) to obtain an equation that we can solve to express the wavenumbers $k$ at which the potential supports semiclassical left-invisibility in terms of $r$, $\varphi_\pm$, $\varepsilon_\pm$, and $m$. Substituting the result in (\ref{condi-ref-1}) and using
(\ref{abs-npm}) and (\ref{polar-npm}), we obtain a complex equation that involves $\delta$. Because $\delta$ is given by the integral of $\fn(\tau)$ over the range of the potential, we need the specific form of $\fn(\tau)$ to determine if this equation has a solution.

As an illustrative example, consider the ansatz:
	\be
	\fn(\tau)=\left\{\begin{array}{cc}
	(kL)^{-1}\left[\fn_-(kL-\tau)\,f(\tau)+ \fn_+\tau \,f(kL-\tau)\right]
		&{\rm for}~\tau\in[0,kL],\\
		1 &{\rm for}~\tau\notin[0,kL],\end{array}\right.
	\label{eg-f}
	\ee
where $f:\R\to\C$ is any piecewise continuous function satisfying $f(0)=1$. Substituting (\ref{eg-f}) in (\ref{fas-2}) and performing a couple of integration by parts give rise to the following remarkably simple relation.
	\be
	\delta=(\fn_-+\fn_+)\cF(kL),
	\label{delta=eg-f}
	\ee
where
	\begin{align*}
	& \cF(\tau):=\frac{1}{kL}\int_0^\tau F(\tau')d\tau',
	&& F(\tau):=\int_0^\tau f(\tau')d\tau'.
	\end{align*}
In light of (\ref{delta=eg-f}), we can write (\ref{condi-ref-1}) in the form
	\be
	\cF(kL)=\frac{1}{2(\fn_-+\fn_+)}\left\{2\pi n+i\ln\left[\frac{(\fn_-+1)(\fn_+-1)}{
	(\fn_--1)(\fn_++1)}\right]\right\},
	\label{condi1-eg-f}
	\ee
where $n\in\Z$ is a mode number.

If we solve (\ref{condi-inv-3}) for $kL$ and substitute the result in (\ref{condi1-eg-f}), we find a complex equation involving $\fn_\pm$. As we described above we can express these in terms of the three real parameters $r$ and $\varphi_\pm$, and the signs $\varepsilon_\pm$. In principle we can use the obtained complex equation to express two of the real parameters in terms of the other parameters. We arrive at the same conclusion by solving (\ref{condi1-eg-f}) for $kL$ in terms of $\fn_\pm$ and inserting the result in (\ref{condi-inv-3}). We implement the latter prescription for the following particularly simple choice for $f$.
	\be
	f(\tau)=1+\frac{\fa\,\tau}{kL},
	\label{f=156}
	\ee
where $\fa$ is a complex number. In view of (\ref{eg-f}), this corresponds to
	\bea
	&&\fn(\tau)=\left\{
    \begin{array}{cc}
    \fn_-+\mbox{\large$\frac{\left[(\fn_+-\fn_-)+\fa(\fn_++\fn_-)\right]\tau}{k L}-
    \frac{\fa(\fn_++\fn_-)\tau^2}{(k L)^2}$}
    &{\rm for}~\tau\in[0,kL],\\[6pt]
    1 & {\rm for}~\tau\notin[0,kL],
    \end{array}\right.
	\label{profile-n}\\[12pt]
	&&v(x)=\left\{
    \begin{array}{cc}
    k^2\left[1-\left\{
	\fn_-+\mbox{\large$\frac{\left[(\fn_+-\fn_-)+\fa(\fn_++\fn_-)\right]x}{L}-
    \frac{\fa(\fn_++\fn_-)x^2}{L^2}$}\right\}^2\right]
    &{\rm for}~x\in[0,L],\\[6pt]
    0 &{\rm for}~\tau\notin[0,L],\end{array}\right.~~~~~~~~
	\\[12pt]
    &&\cF(kL)= \frac{(3+\fa)kL}{6},
    \label{ez923}
	\eea
and (\ref{condi1-eg-f}) takes the form
	\be
	kL=\frac{3}{(3+\fa)(\fn_-+\fn_+)}\left\{2\pi n+i\ln\left[\frac{(\fn_-+1)(\fn_+-1)}{
	(\fn_--1)(\fn_++1)}\right]\right\}.
	\ee
Now, we can substitute this equation in (\ref{condi-inv-3}) express $\fn_\pm$ in terms of $r,\varphi_\pm$ and $\varepsilon_\pm$ in the resulting equation and use it to fix two of the variable $r,\varphi_\pm$ in terms of the other and $\varepsilon_\pm$. Because this equation is a highly complicated transcendental equation, we can only do this provided that we know the physically relevant ranges of the parameters of the problem. In the next section, we focus our attention to applications in optics where the physically acceptable ranges of the parameters are well-known. In the remainder of this section, we comment on the simplifications occurring for $\cP\cT$-symmetric potentials for which
    \be
    \fn(kL-\tau)^*=\fn(\tau).
    \label{PT-sym}
    \ee

As explained in Ref.~\cite{pra-2013a} the phenomenon of unidirectional invisibility is fundamentally $\cP\cT$-symmetric in the sense that the $\cP\cT$-reflection transformation leaves the equations characterizing unidirectionally invisible configurations invariant. As a byproduct of this symmetry, the $\cP\cT$-symmetric unidirectionally invisible configurations have a much simpler structure. This behavior extends to semiclassical unidirectional invisibility.

For a $\cP\cT$-symmetric finite-range potential satisfying (\ref{PT-sym}), we have $\fn_-=\fn_+^*$ (so that $\eta_-=\eta_+$ and $\kappa_-=-\kappa_+$) and $\delta=\int_0^{kL}
\eta(\tau)d\tau$, where $\eta(\tau)$ is the real part of $\fn(\tau)$, and $\eta_\pm$ and $\kappa_\pm$ are respectively the real and imaginary parts of $\fn_\pm$. In light of these relations, Eqs.~(\ref{condi-ref-1}) and (\ref{condi-inv-1}) respectively take the form
    \bea
    &&\int_0^{kL}\eta(\tau)d\tau=\pi m_1-\tan^{-1} \left(\frac{2\kappa_+}{\eta_+^2-1+\kappa_+^2}\right),
    \label{condi-ref-1-pt}\\
    &&kL=\pi m_2-\tan^{-1}\left[\frac{(\eta_+^2+1)\kappa_+}{(\eta^2_+-1)\eta_+}\right],
    \label{condi-inv-1-pt}
    \eea
where $m_1$ and $m_2$ are integers. Therefore, the condition that the potential supports semiclassical left-reflectionlessness slightly restricts the form of the real part of $\fn(\tau)$ while demanding semiclassical left-invisibility fixes the wavelengths at which this phenomenon occurs. In other words, for $\cP\cT$-symmetric potentials the equations characterizing semiclassical unidirectional invisibility decouple.

For the class of potentials satisfying~(\ref{eg-f}) the $\cP\cT$-symmetry~(\ref{PT-sym}) is equivalent to demanding that $f$ be a real-valued function and $\fn_-=\fn_+^*$. In particular, for functions $f$ of the form (\ref{f=156}), the former condition amounts to choosing $\fa$ to be a real parameter. In this case, imposing the condition that the potential displays semiclassical left-reflectionlessness (\ref{condi-ref-1-pt}) determines the allowed values of $\fa$ according to
    \be
    \fa=\frac{3}{\eta_+kL}\left[\pi m_1-\tan^{-1} \left(\frac{2\kappa_+}{\eta_+^2-1+\kappa_+^2}\right)\right]-3.
    \label{condi-ref-1-pt-2}
    \ee

\section{Optical Applications}

In effectively one-dimensional optical setups the Helmholtz equation, which describes the propagation of time-harmonic electromagnetic waves, takes the form of the time-independent Schr\"odinger equation (\ref{sch}), and $\fn$ plays the role of the complex refractive index of the medium in which the electromagnetic waves propagate. Typically, the real part of $\fn$ takes positive values of the order of 1 and its imaginary part is at least three orders of magnitude smaller than its real part. Therefore, recalling that we denote the real and imaginary parts of $\fn_\pm$ respectively by $\eta_\pm$ and $\kappa_\pm$, so that
$\fn_\pm=\eta_\pm+i\kappa_\pm$, we have $\kappa_\pm\ll \eta_\pm$ and consequently
    \be
    \varphi_\pm:=\tan^{-1}(\mbox{\large$\frac{\kappa_\pm}{\eta_\pm}$})\approx \mbox{\large$\frac{\kappa_\pm}{\eta_\pm}$}\ll 1.
    \label{phi-approx}
    \ee
Using this relation in (\ref{abs-npm}) yields
    \bea
    \eta_\pm\approx \eta:=\sqrt{1+\frac{r^2}{2}\left(1+\sqrt{1+\frac{4}{r^2}}\right)}>1,
    \label{eta-pm=}
    \eea
where $\approx$ stands for the fact that we ignore the quadratic and higher order terms in powers of $\kappa_\pm$. In particular, for all practical purposes, we can take $\eta_+=\eta_-=\eta$.

Next, we introduce the parameters
    \begin{align*}
    &\chi_\pm:=\frac{\kappa_\pm}{\eta^2-1}, &&
    \xi:=2(\chi_--\chi_+)=\frac{2(\kappa_--\kappa_+)}{\eta^2-1},
    \end{align*}
and make use of (\ref{phi-approx}) and (\ref{eta-pm=}) to express (\ref{condi-ref-1}) and (\ref{condi-inv-3}) in the form
    \bea
    && e^{2i\delta}\approx 1+i\,\xi ,
    \label{ref-2n}\\
    && kL\approx \pi m+ \frac{(\eta^2+1)\xi}{4\eta },
    \label{eqz1}
    \eea
respectively. Applying (\ref{phi-approx}) once again in (\ref{ref-2n}), we have
    \be
    \delta\approx \pi n +\frac{1}{2}\tan^{-1}\xi-\frac{i}{4}\ln(1+\xi^2),
    \label{ref-3n}
    \ee
where $n$ is an integer. Similarly, employing (\ref{phi-approx}) and (\ref{eta-pm=}) in
(\ref{RR-sc-ref}), we find
    \be
    |R^r|\approx
    |\kappa_--\kappa_+| ~\sqrt{\frac{\eta^{-2}+(\chi_++\chi_-)^2}{\eta^{-2}+(2\chi_++\chi_-)^2}}=
    \frac{(\eta^2-1)|\xi|}{2}~\sqrt{\frac{\eta^{-2}+(\chi_++\chi_-)^2}{\eta^{-2}+(2\chi_++\chi_-)^2}}.
    \label{Rr-app-13}
    \ee

Relations~(\ref{eqz1}), (\ref{ref-3n}), and (\ref{Rr-app-13}), with $\eta>1$ and $\xi\neq 0$, describe the phenomenon of semiclassical left-invisibility. As seen from (\ref{Rr-app-13}), $|R^r|$ is of the same order of magnitude as $|\kappa_\pm|$, unless $\kappa_-\approx -2\kappa_+$ and $\eta-1\ll |\kappa_+|$. In typical situations, $\eta-1$ is of the same order of magnitude as $|\kappa_\pm|$, $\xi$ is of order $1$, and $|R^r|$ takes relatively small values. Notice that because $kL\gg 1$, according to (\ref{eqz1}), $m$ is a large positive integer.

As a concrete example, consider the refractive index profiles of the form (\ref{eg-f}). Then $\delta$ is given by (\ref{delta=eg-f}) and in view of (\ref{ref-3n}) and (\ref{phi-approx}), we have
    \be
    \cF(kL)\approx\frac{ \pi n}{2\eta} +\frac{1}{4\eta}\tan^{-1}\xi-
    \frac{i}{8\eta}\ln(1+\xi^2).
    \label{FkL=101}
    \ee
For a quadratic refractive index of the form (\ref{profile-n}), we can use (\ref{ez923}) to express (\ref{FkL=101}) as
    \be
    kL\approx\frac{4\pi n+2\tan^{-1}\xi-i\ln(1+\xi^2)}{4(1+\fa/3)\eta}.
    \label{quad-1}
    \ee
Eliminating $kL$ from this equation and (\ref{eqz1}), we find\footnote{Notice that $\fa$ determines the shape of the refractive index.}
    \be
    \fa\approx 3\left[\frac{4\pi n+2\tan^{-1}\xi-i\ln(1+\xi^2)}{4\pi m\eta+(\eta^2+1)\xi}-1\right].
    \label{a=178}
    \ee
Expressing $\fn$ as a function $\hat x:=x/L$ and using (\ref{profile-n}) and (\ref{eta-pm=}), we have
	\be
	\fn \approx
	\left\{\begin{array}{cc}
	\eta+i\kappa_-+\left[2\fa(\eta+i\bar\kappa)+i\kappa\right]\hat x-
	2\fa(\eta+i\bar\kappa)\hat x^2 & \mbox{for}~\hat x\in[0,1],\\[6pt]
	1 & \mbox{for}~\hat x\notin[0,1],\end{array}\right.
	\label{n-profile-21}
	\ee
where $\kappa:=\kappa_+-\kappa_-$ and $\bar\kappa:=(\kappa_++\kappa_-)/2$. According to (\ref{phi-approx}), (\ref{eta-pm=}), and (\ref{n-profile-21}), $|\fn|$ is of the same order of magnitude as $|\fn_\pm|$ or $\eta$ provided that $|\fa|$ is at most of order $1$. In view of this observation, the fact that $\eta$ and $\xi$ are of order 1 and $m\gg 1$, we can use (\ref{a=178}) to infer that
    \begin{align}
    \fa\approx 3\left(\frac{n}{m\eta}-1\right).
    \label{a=174}
    \end{align}
Imposing the condition $|\fa|\lessapprox 1$, we then obtain
    \be
    \frac{2m}{3}\lessapprox n \lessapprox \frac{4m}{3}.
    \label{a=185}
    \ee
We also note that for the choice $m=n$, $\fa\approx 3(\eta^{-1}-1)\approx 1-\eta$, where we have used the fact that $\eta-1\ll 1$.

For definiteness, let us consider a sample with $\eta=1.001$, $\kappa_-=-0.002$, and $\kappa_+=0.001$, so that $\xi=-2.999$. Then according to (\ref{eqz1}) and (\ref{Rr-app-13}), $kL\approx\pi m-1.499$ and $|R^r|\approx 0.003354$. With $L=100~\mu{\rm m}$ and $\lambda=2\pi/k$ in the range $100~{\rm nm}$ to $2~\mu{\rm m}$, we have $100\pi \lessapprox kL\lessapprox 2000\pi$. This implies
    \be
    100\lessapprox m \lessapprox 2000.
    \label{range-m}
    \ee
It is not difficult to show that $m\gtrapprox 100$ is a sufficient condition for the validity of the semiclassical approximation. To see this, we note that the latter is equivalent to the adiabaticity condition (\ref{adi-condi-0}), which we can also express as
    \be
    \left|\frac{1}{\fn}\frac{d\fn}{d\hat x}\right|\ll 4 k L~~~~~{\rm for}~~~~0\leq\hat x\leq 1.
    \label{sc-condi-376}
    \ee
Because $\eta$ and $|\fa|$ are of order $1$ and $|\kappa_\pm|\ll 1<\eta$, $\left|\frac{1}{\fn}\frac{d\fn}{d\hat x}\right|$ is also of order $1$. This shows that for $m\geq 100$, which corresponds to $kL> 300$, (\ref{sc-condi-376}) holds and the semiclassical approximation is reliable.

A closer examination of (\ref{n-profile-21}) reveals the fact that $|\RE(\fn)|$ is very sensitive to the value of $|m-n|$. For example for $m=300$ increasing $n$ from 300 to 301 results in an increase in the maximum value of $\RE(\fn)-\eta$ by about two orders of magnitude. This makes the choice $m=n$ more relevant for an experimental investigation of the system.

\begin{figure}[t]
	\begin{center}
	\includegraphics[scale=.6]{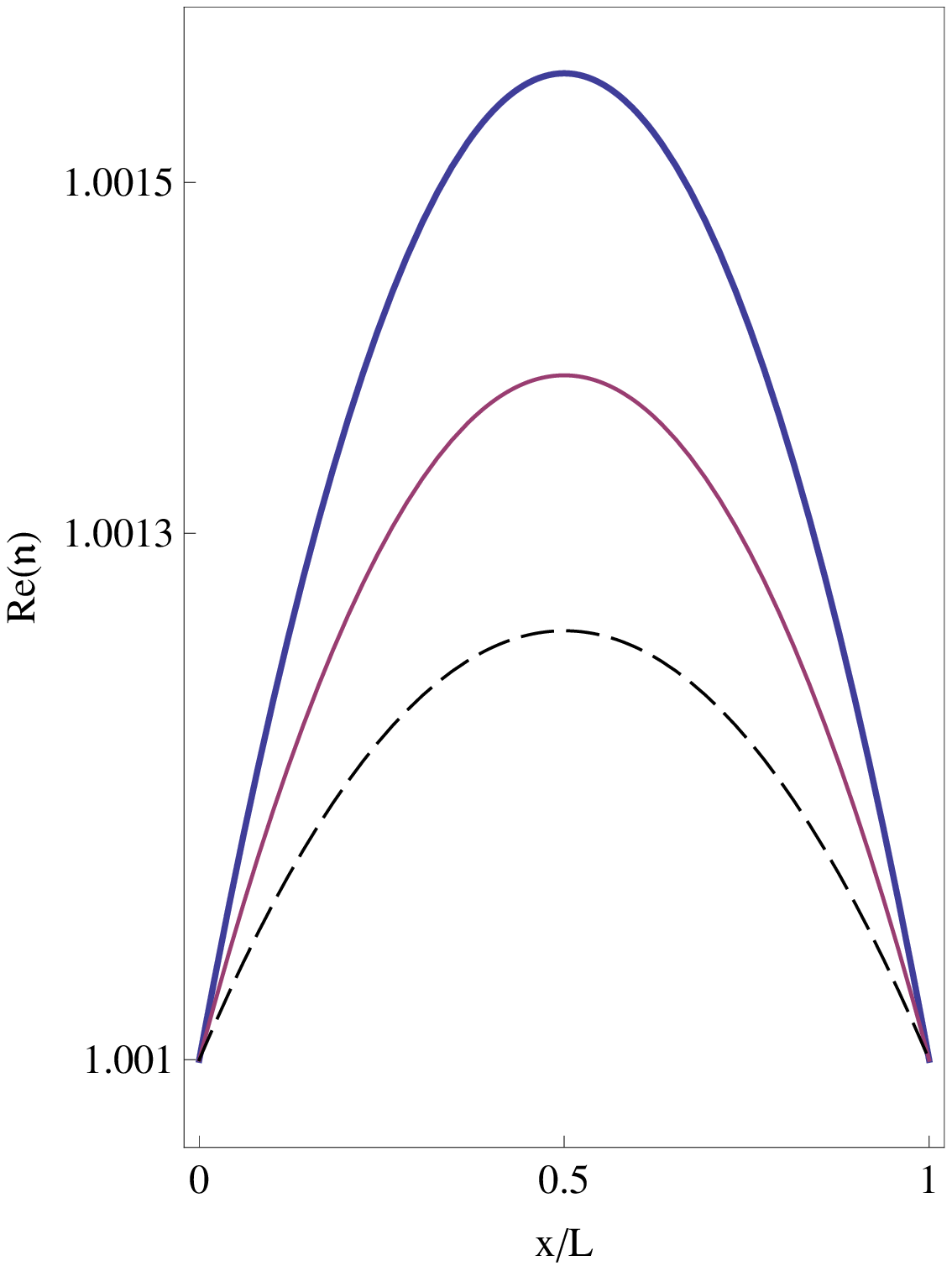}\hspace{1.8cm}\includegraphics[scale=.6]{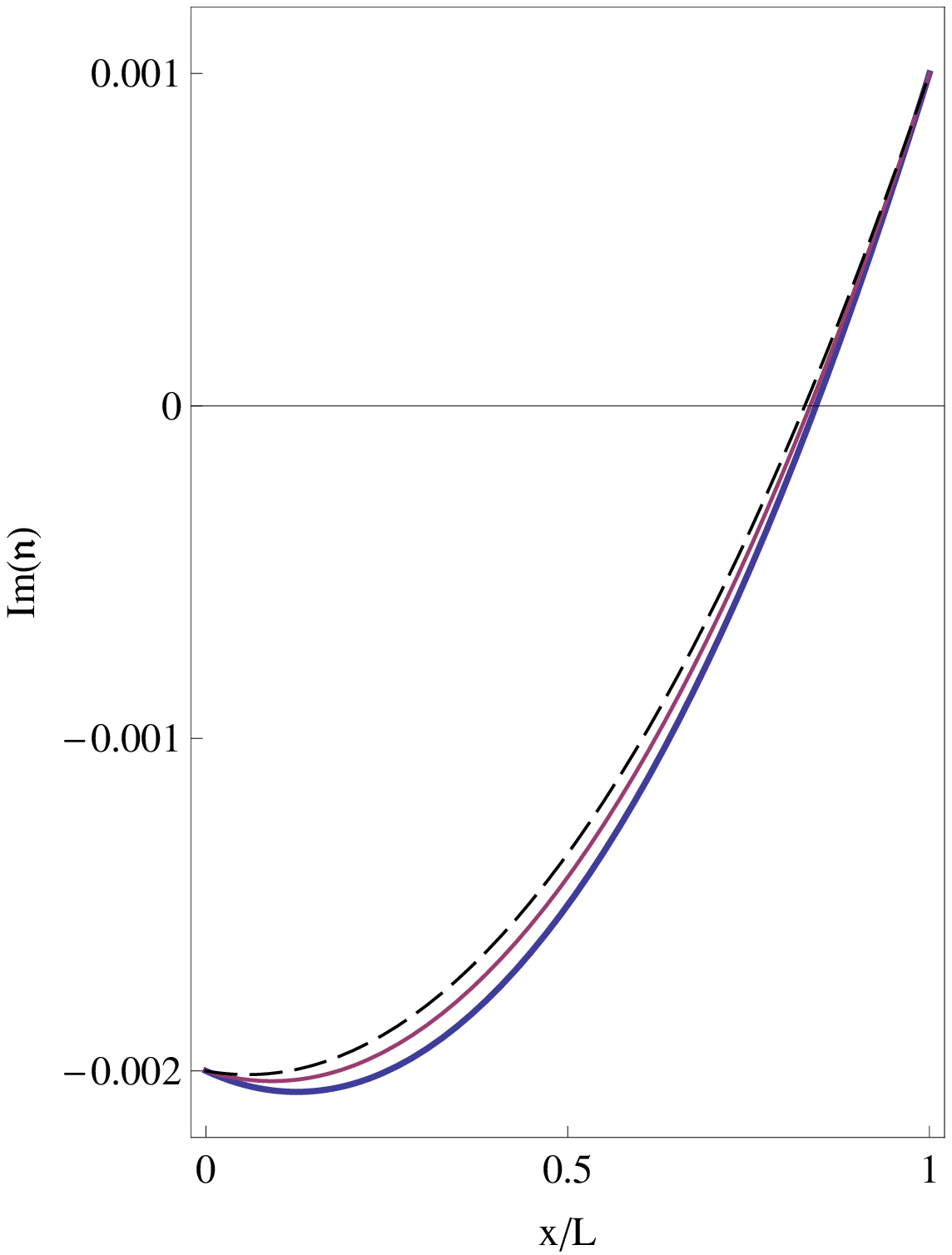}
	\caption{(Color online) Graphs of the real (on the left) and imaginary (on the right) parts of the refractive index (\ref{n-profile-21}) as a function of $x/L$ for $\eta=1.001$, $\kappa_-=-0.002$, and $\kappa_+=0.001$. The thick (dark blue), thin (purple), and the dashed (black) curves correspond to $m=n=275,300,$ and 325, respectively. For $L=100~\mu{\rm m}$ these configurations are unidirectional invisibility from the left for $\lambda=728.537,\,667.729$, and $616.290$~nm, respectively. The portion of the $\IM(\fn)$ curves that lies below (above) the horizonal $\IM(\fn)=0$ line corresponds to the gain (lossy) regions.}
	\label{fig1}
	\end{center}
	\end{figure}

Figure~\ref{fig1} shows the graphs of the real and imaginary parts of the refractive index~(\ref{n-profile-21}) for $m=n=275,300,$ and $325$. These correspond to situations where the system is left-invisible, with $|R^r|\approx 0.0034$, for $\lambda=728.537,\,667.729$, and $616.290$~nm, respectively.

Table~\ref{table1} gives the values of the physical parameters that makes the optical system described by (\ref{n-profile-21}) left-invisible for various values $m$.
    \begin{table}
    \begin{center}
    \begin{tabular}{|c|c|c|c|}
    \hline
    $m$ & $kL$ & $\lambda$ (nm) & $\fa$ \\
    \hline
    100 & $312.660$ & 2009.59 &$(8.371-5.515 i)\times 10^{-3}$\\
    $\vdots$ & $\vdots$ & $\vdots$ &$\vdots$\\
    298 & $934.695$ & 672.217 &$(8.057-18.450 i)\times 10^{-4}$\\
    299 & $937.837$ & 669.966 &$(7.929-18.389 i)\times 10^{-4}$\\
    300 & $940.979$ & 667.729 &$(7.802-18.327 i)\times 10^{-4}$\\
    301 & $944.120$ & 666.507 &$(7.677-18.266 i)\times 10^{-4}$\\
    302 & $947.262$ & 663.300 &$(7.552-18.206 i)\times 10^{-4}$\\
    $\vdots$ & $\vdots$ & $\vdots$ &$\vdots$\\
    2000 & $6281.690$ & 100.024 &$(-2.431-0.275 i)\times 10^{-3}$\\
    \hline
    \end{tabular}
    \vspace{6pt}
    \caption{Values of $m$, $k L$, $\lambda$, and $\fa$ for which the optical system with refractive index (\ref{n-profile-21}) displays left-invisibility for $\eta=1.001$, $\kappa_-=-0.002$, $\kappa_+=0.001$, $L=100~\mu{\rm m}$, and $n=m$.}
    \label{table1}
    \end{center}
    \end{table}
The numerical values of $\fa$ confirm our expectation that for all values of $m$ in the range (\ref{range-m}) we have $kL\gtrapprox 300$ and $|\fa|\lessapprox 10^{-3}$. The latter implies that $|\fn-1|\lessapprox
10^{-3}$. Therefore, perturbation theory provides another reliable method of solving the scattering problem for this model.

As we discussed in Section~4, the construction of $\cP\cT$-symmetric potentials displaying semiclassical unidirectional invisibility is more straightforward. For example for the $\cP\cT$-symmetric models of the form (\ref{n-profile-21}), where $\kappa_-=-\kappa_+$ and $\fa$ is real, we can determine the values of $kL$ and $\fa$ that support semiclassical left-invisibility by
choosing and inserting the values of $\eta_+,\kappa_+,m_1$, and $m_2$ in Eqs.~(\ref{condi-inv-1-pt}) and (\ref{condi-ref-1-pt-2}), respectively. An interesting example, is $\eta_+=1.01$, $\kappa_+=0.001$, $m_1=303$, and $m_2=300$, where in the optically active region the real part of $\fn$ is essentially constant while its imaginary part is a linear function of $x$; $\fn\approx 1.01-0.001 i(1-2x/L)$ for $x\in[0,L]$. For this configuration $kL=942.379$, $\fa=3.16\times 10^{-6}$, and for $L=100~\mu{\rm m}$ we have $\lambda=666.737~{\rm nm}$.

\section{Concluding Remarks}

The formulation of the scattering problem in terms of a time-dependent Schr\"odinger equation with a time-dependent Hamiltonian operator suggests the use of the adiabatic approximation in the study of the quantum potential scattering. This turns out to coincide with the application of the semiclassical approximation in scattering theory. It is remarkable that the geometric part of the complex phase of the evolving state vectors gives rise to the pre-exponential factor in the WKB wave functions. This provides another intriguing manifestation of the role of geometric phases in quantum mechanics.

In this article we have used adiabatic approximation to derive a semiclassical expression for the transfer matrix of a general finite-range potential that can be complex or even energy-dependent. We have then employed this expression in the study of the phenomenon of unidirectional invisibility. In particular, we have introduced the notions of semiclassical unidirectional reflectionlessness and invisibility and established the fact that the reflection and transmission amplitudes take the same values at all the wavelengths for which the potential displays semiclassical unidirectional reflectionlessness. We have also offered a detailed examination of the optical realizations of semiclassical unidirectional invisibility and constructed concrete optical potentials possessing this property.

As pointed out by one of the  referees, the connection between semiclassical and adiabatic approximations that is revealed in this article raises the possibility of the application of the results obtained within the context of ``shortcuts to adiabaticity" \cite{SA} in one-dimensional scattering theory.

\subsection*{Acknowledgments}  This work has been supported by  the Scientific and Technological Research Council of Turkey (T\"UB\.{I}TAK) in the framework of the project no: 112T951, and by the Turkish Academy of Sciences (T\"UBA).

\np

\ed